\documentclass[graybox]{svmult}

\usepackage{mathptmx}       
\usepackage{helvet}         
\usepackage{courier}        
\usepackage{type1cm}        
\usepackage{makeidx}         
\usepackage{graphicx}        
\usepackage{multicol}        
\usepackage[bottom]{footmisc}

\makeindex             

\begin{document}

\title*{The Frontier of Reionization: Theory and Forthcoming Observations}
\titlerunning{Reionization} 
\author{Abraham Loeb}
\institute{Harvard University, CfA, MS 51, 60 Garden Street, Cambridge MA
02138, {aloeb@cfa.harvard.edu} }
%
%
\maketitle 

\abstract{The cosmic microwave background provides an image of the Universe
0.4 million years after the Big Bang, when atomic hydrogen formed out of
free electrons and protons.  One of the primary goals of observational
cosmology is to obtain follow-up images of the Universe during the epoch of
reionization, hundreds of millions of years later, when cosmic hydrogen was
ionized once again by the UV photons emitted from the first galaxies.  To
achieve this goal, new observatories are being constructed, including
low-frequency radio arrays capable of mapping cosmic hydrogen through its
redshifted 21cm emission, as well as imagers of the first galaxies such as
the {\it James Webb Space Telescope (JWST)} and large aperture ground-based
telescopes.  The construction of these observatories is being motivated by
a rapidly growing body of theoretical work.  Numerical simulations of
reionization are starting to achieve the dynamical range required to
resolve galactic sources across the scale of hundreds of comoving Mpc,
larger than the biggest ionized regions.  }

\section{Preface}
\label{sec:1}

When we look at our image reflected off a mirror at a distance of 1 meter,
we see the way we looked 6.7 nanoseconds ago, the light travel time to the
mirror and back. If the mirror is spaced $10^{19}~{\rm cm} \simeq 3~$pc
away, we will see the way we looked twenty-one years ago. Light propagates
at a finite speed, and so by observing distant regions, we are able to see
what the Universe looked like in the past, a light travel time ago. The
statistical homogeneity of the Universe on large scales guarantees that
what we see far away is a fair statistical representation of the conditions
that were present in our region of the Universe a long time ago.  This
fortunate situation makes cosmology an empirical science. We do not need to
guess how the Universe evolved. Using telescopes we can simply see how the
Universe appeared at earlier cosmic times. In principle, this allows the
entire 13.7 billion year cosmic history of our Universe to be reconstructed
by surveying galaxies and other sources of light out to large distances.
From these great distances, the wavelength of the emitted radiation is
stretched by a large redshift factor $(1+z)$ until it is observed, owing to
the expansion of the Universe. Since a greater distance means a fainter
flux from a source of a fixed luminosity, the observation of the earliest,
highest-redshift sources of light requires the development of sensitive
infrared telescopes such as the {\it James Webb Space Telescope (JWST)}.

Our cosmic photo album contains an early image of the Universe when it was
0.4 million years old in the form of the cosmic microwave background (CMB)
\cite{Bennett,WMAP}, as well as many snapshots of galaxies more than a
billion years later ($z<6$; see overview in Ref. \cite{Ellis}).  But we are
still missing some crucial pages in this album.  In between these two
epochs was a period when the Universe was dark, stars had not yet formed,
and the cosmic microwave background no longer traced the distribution of
matter.  And this is precisely the most interesting period, when the
primordial soup evolved into the rich zoo of objects we now see.  The
situation that cosmologists face is similar to having a photo album of a
person that contains the first ultrasound image of him or her as an unborn
baby and some additional photos as a teenager and an adult.  If you tried
to guess from these pictures what happened in the interim, you could be
seriously wrong. A child is not simply a scaled-up fetus or scaled-down
adult. The same is true with galaxies. They did not follow a
straightforward path of development from the incipient matter clumping
evident in the microwave background.

\section{Preliminaries}

About $400,000$ years after the Big Bang the temperature of the Universe
dipped for the first time below a few thousand degrees Kelvin. The protons
and electrons were then sufficiently cold to recombine into hydrogen atoms.
It was just before the moment of cosmic recombination (when matter started
to dominate in energy density over radiation) that gravity started to
amplify the tiny fluctuations in temperature and density observed in the
CMB data \cite{WMAP}. Regions that started out slightly denser than average
began to contract because the gravitational forces were also slightly
stronger than average in these regions. Eventually, after hundreds of
millions of years of contraction, galaxies and the stars within them were
able to form.

\begin{figure}
\centering
\includegraphics[height=6cm]{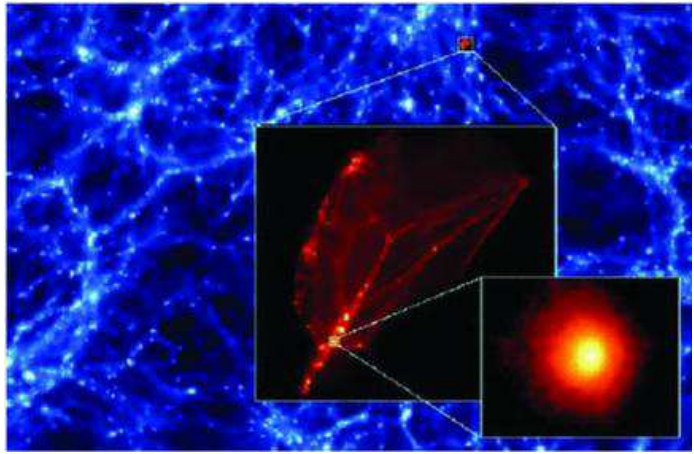}
\label{figMoore}
\caption{A slice through a numerical simulation of the first dark matter
condensations to form in the Universe (from Diemand, Moore, \& Stadel
2005). Colors represent the dark matter density at $z=26$. The simulated
volume is 60 comoving pc on a side, simulated with 64 million particles
each weighing $1.2\times 10^{-10}M_\odot$. }
\end{figure}

The detailed statistical properties of the CMB anisotropies \cite{WMAP}
indicate that indeed the structure apparent in the present-day Universe was
seeded by small-amplitude inhomogeneities, mostly likely induced by quantum
fluctuations during the early epoch of inflation.  The growth of
structure from these seeds was enhanced by the presence of dark matter --
an unknown substance that makes up the vast majority (84\%) of the cosmic
density of matter.  The motion of stars and gas around the centers of
nearby galaxies indicates that each is surrounded by an extended mass of
dark matter, and so dynamically-relaxed dark matter concentrations are
generally referred to as ``halos''.

Under the assumption that general relativity describes the evolution of
the Universe, the measured CMB anisotropies indicate conclusively that most
of the matter in the Universe must be very weakly coupled to
electromagnetism and hence cannot be the matter that we are made of
(baryons).  This follows from the fact that prior to hydrogen
recombination, the cosmic plasma was coupled to the radiation through
Thomson scattering.  Small-scale fluctuations were then damped in the
radiation-baryon fluid by photon diffusion. The damping is apparent in the
observed suppression of the CMB anisotropies on angular scales well below a
degree on the sky, corresponding to spatial scales much smaller than 200
comoving Mpc.  To put this scale in context, the matter that makes up
galaxies was assembled from scales of $<2$Mpc.  In order to preserve the
primordial inhomogeneities that seeded the formation of galaxies, it is
necessary to have a dominant matter component that does not couple to the
radiation fluid. The most popular candidate for making up this component is
a weakly interacting massive particle (WIMP). If this particle is the
lightest supersymmetric particle, it might be discovered over the coming
decade in the data stream from the {\it Large Hadron Collider}.

The natural temperature for the decoupling of WIMPs is expected to be high
(tens of MeV), allowing them to cool to an extremely low temperature by the
present epoch. The resulting {\it Cold Dark Matter (CDM)} is expected to
fragment down to a Jupiter mass scale \cite{Moore,LZ,Bert}.  The baryons,
however, cannot follow the CDM on small scales because of their higher
thermal pressure. The minimum scale for the fragmentation of the baryons,
the so-called ``filtering scale'' (which is a time-averaged Jeans mass),
corresponds to $\sim 10^5M_\odot$ prior to reionization \cite{SAA}.

\subsection{The First Stars}

For the scale-invariant $\Lambda$CDM power spectrum \cite{WMAP}, the first
dark matter halos to contain gas have formed at a redshift of several tens.
The assembly and cooling of gas in these halos resulted in the formation of
the first stars \cite{Bromm}.  Hydrodynamical simulations indicate that the
primordial (metal-free) gas cooled via the radiative transitions of
molecules such as H$_2$ and HD down to a temperature floor of a few hundred
K, dictated by the energy levels of these molecules.  At the characteristic
density interior to the host clouds, the gas fragmented generically into
massive ($>100M_\odot$) clumps which served as the progenitors of the first
stars. The relatively high sound speed ($c_s$) resulted in a high accretion
rate ($\dot{M}\sim c_s^3/G$ over the stellar lifetime of a few million
years) and a high characteristic mass for the first (so-called {\it
Population III}) stars \cite{BLo,Abel}.  The lowest-mass halos most likely
hosted one star per halo.

Population III stars in the mass range of 140--260$M_\odot$ led to
pair-instability supernovae that enriched the surrounding gas with heavy
elements \cite{Heger}.  Enrichment of the gas to a carbon or oxygen
abundance beyond $\sim 10^{-3.5}Z_\odot$ resulted in efficient cooling and
fragmentation of the gas to lower-mass stars \cite{Met,Frebel,Schneider}.
The hierarchical growth in halo mass eventually led to the formation of
halos with a virial temperature of $\sim 10^4$K in which cooling was
mediated by atomic transitions. Fragmentation of gas in these halos could
have led to the direct formation of the seeds for quasar black holes
\cite{BH,Rasio}.

\begin{figure}
  \includegraphics[height=.3\textheight]{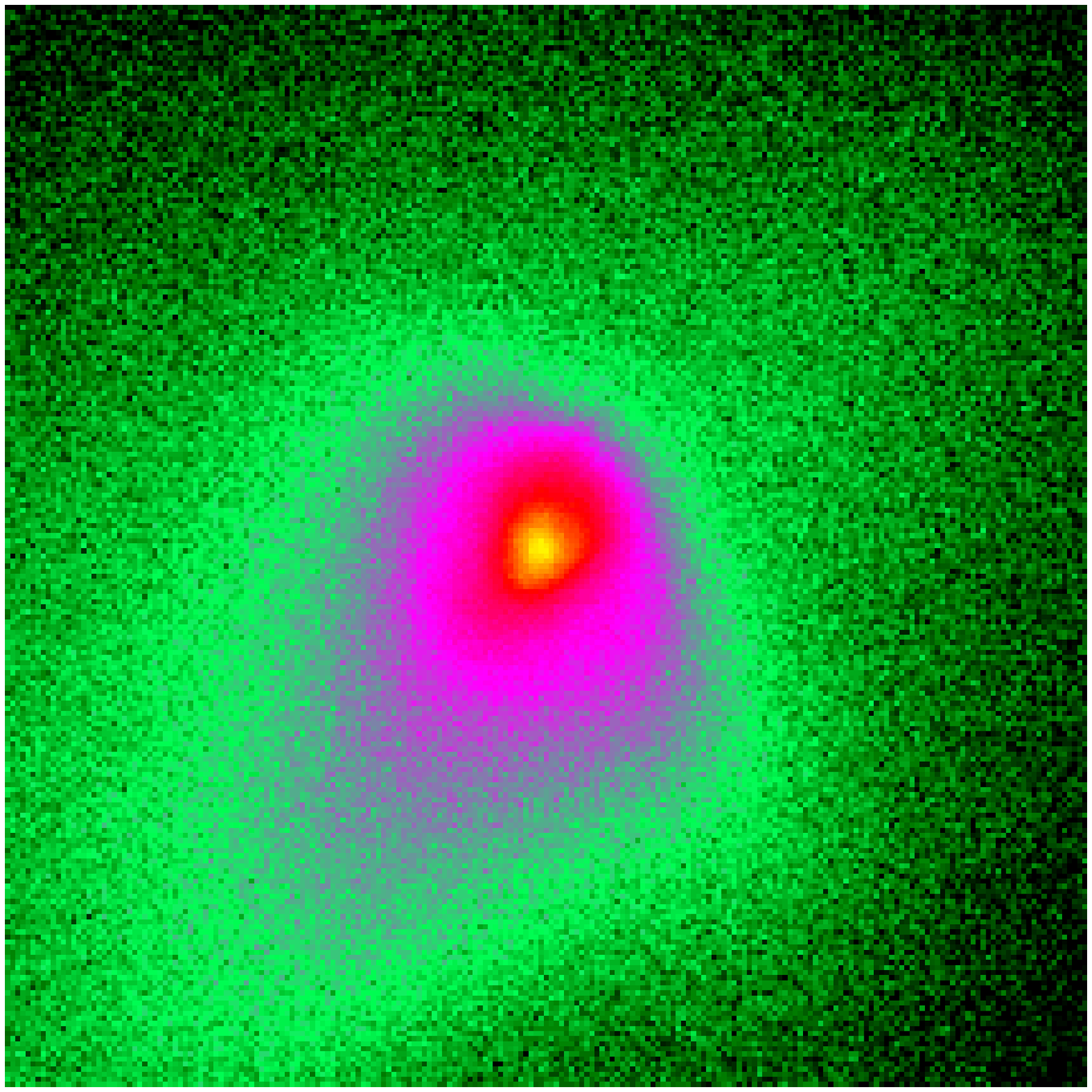}
  \includegraphics[height=.3\textheight]{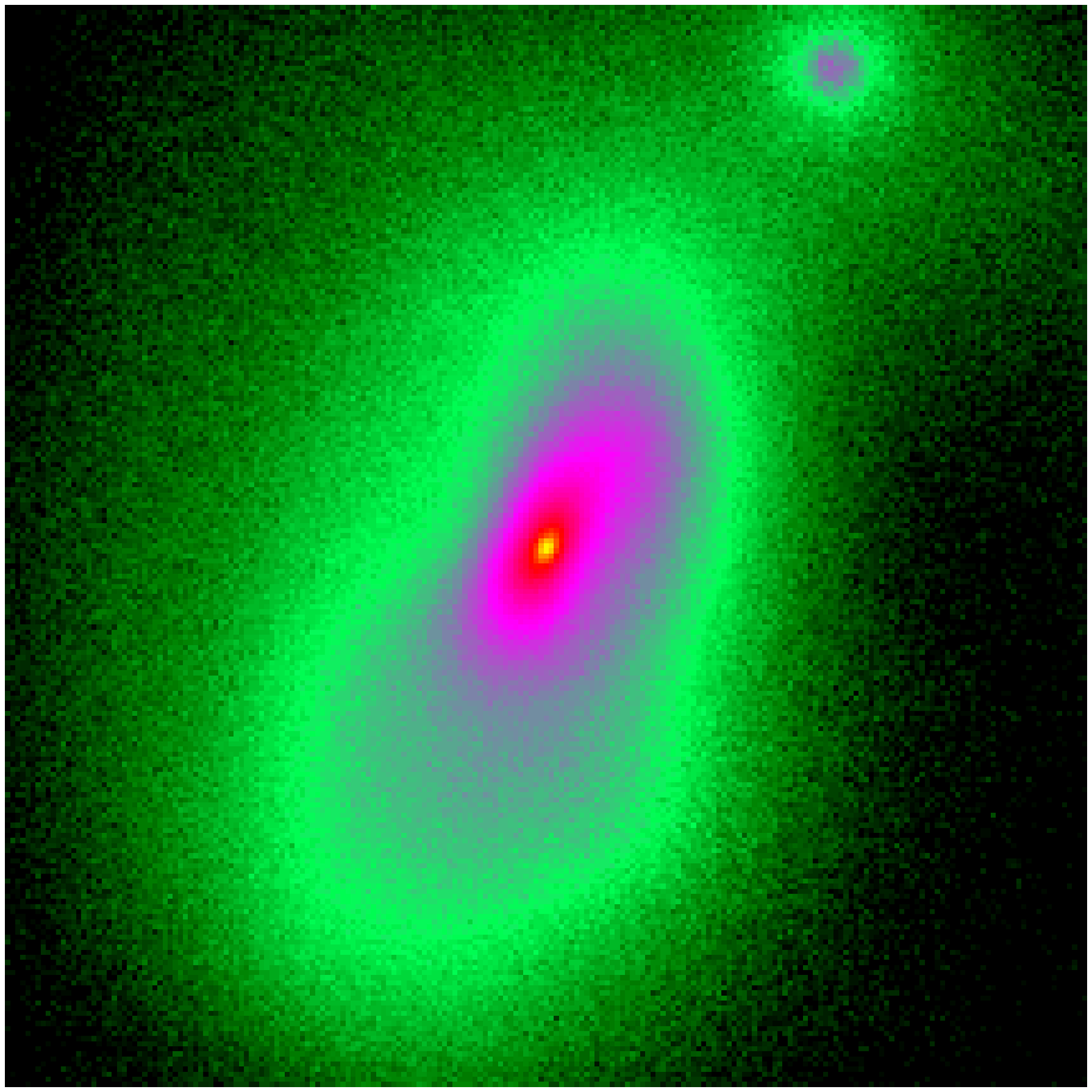}
\caption{Collapse and fragmentation of a primordial cloud (from Bromm \&
Loeb 2004).  Shown is the projected gas density at a redshift $z\simeq
21.5$, briefly after gravitational runaway collapse has commenced in the
center of the cloud.  {\it Left:} The coarse-grained morphology in a box
with linear physical size of 23.5~pc.  At this time in the unrefined
simulation, a high-density clump (sink particle) has formed with an initial
mass of $\sim 10^{3}M_{\odot}$.  {\it Right:} The refined morphology in a
box with linear physical size of 0.5~pc.  The central density peak,
vigorously gaining mass by accretion, is accompanied by a secondary clump.}
\label{2ab}
\end{figure}

A massive metal-free star is an efficient factory for the production of
ionizing photons. Its surface temperature ($\sim 10^5$K) and emission
spectrum per unit mass are nearly independent of its mass above a few
hundred $M_\odot$, as it radiates at the Eddington luminosity (i.e. its
luminosity is proportional to its mass).  Therefore the cumulative
emissivity of the first massive stars was proportional to their total
cumulative mass, independent of their initial mass function. These stars
produced $\sim 10^5$ ionizing photons per baryon incorporated into them
\cite{BKL,TS}. In comparison, low-mass stars produce $\sim 4,000$ ionizing
photons per baryon \cite{SAA}.  In both cases, it is clear that only a small
fraction of the baryons in the Universe needs to be converted into stars in
order for them to ionize the rest.

Given the formation rate of galaxy halos as a function of cosmic time, the
course of reionization can be determined by counting photons from all
sources of light \cite{Arons,Shapiro,Tegmark,Kamionkowski,Fukugita,Shap2,
HL97}. Both stars and black holes contribute ionizing photons, but the
early Universe is dominated by small galaxies which in the local Universe
have central black holes that are disproportionately small, and indeed
quasars are rare above redshift 6 \cite{Fan}. Thus, stars most likely
dominated the production of ionizing UV photons during the reionization
epoch [although high-redshift galaxies should have also emitted X-rays from
accreting black holes and accelerated particles in collisionless shocks
\cite{Oh,FL}]. Since most stellar ionizing photons are only slightly more
energetic than the 13.6 eV ionization threshold of hydrogen, they are
absorbed efficiently once they reach a region with substantial neutral
hydrogen.  This makes the intergalactic medium (IGM) during reionization a
two-phase medium characterized by highly ionized regions separated from
neutral regions by sharp ionization fronts (see Figure~\ref{fig:rei}).

We can obtain a first estimate of the requirements of reionization by
demanding one stellar ionizing photon for each hydrogen atom in the
IGM. If we conservatively assume that stars within the ionizing galaxies
were similar to those observed locally, then each star produced $\sim
4000$ ionizing photons per baryon. Star formation is observed today to
be an inefficient process, but even if stars in galaxies formed out of
only $\sim10\%$ of the available gas, it was still sufficient to
accumulate a small fraction (of order $0.1\%$) of the total baryonic
mass in the Universe into galaxies in order to ionize the entire
IGM. More accurate estimates of the actual required fraction account
for the formation of some primordial stars (which were massive,
efficient ionizers, as discussed above), and for recombinations of
hydrogen atoms at high redshifts and in dense regions.

From studies of quasar absorption lines at $z\sim 6$ we know that the IGM
is highly ionized a billion years after the Big Bang. There are hints,
however, that some large neutral hydrogen regions persist at these early
times \cite{WL04a,Mes,Lidz} and so this suggests that we may not need to go
to much higher redshifts to begin to see the epoch of reionization.  We now
know that the Universe could not have fully reionized earlier than an age
of $\sim 300$ million years, since WMAP3 observed the effect of the freshly
created plasma at reionization on the large-scale polarization anisotropies
of the CMB and this limits the reionization redshift \cite{WMAP}; an
earlier reionization, when the Universe was denser, would have created a
stronger scattering signature that would be inconsistent with the WMAP3
observations. In any case, the redshift at which reionization ended only
constrains the overall cosmic efficiency of ionizing photon production. In
comparison, a detailed picture of reionization as it happens will teach us
a great deal about the population of young galaxies that produced this
cosmic phase transition.

\begin{figure}
\centering
\includegraphics[width=4.6in]{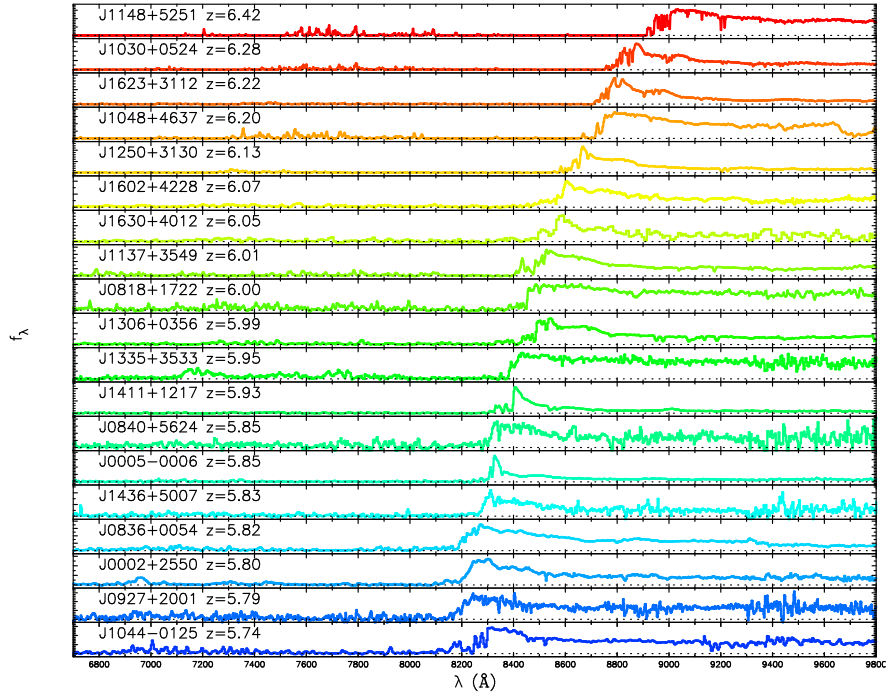}
\caption{Spectra of 19 quasars with redshifts $5.74<z<6.42$ from the
{\it Sloan Digital Sky Survey}, taken from Fan et al. (2005). For some
of the highest-redshift quasars, the spectrum shows no transmitted
flux shortward of the Ly$\alpha$ wavelength at the quasar redshift
(the so-called ``Gunn-Peterson trough''), indicating a non-negligible
neutral fraction in the IGM.}
\label{fig:19qsos}
\end{figure}

\begin{figure}
\centering
\includegraphics[width=2in]{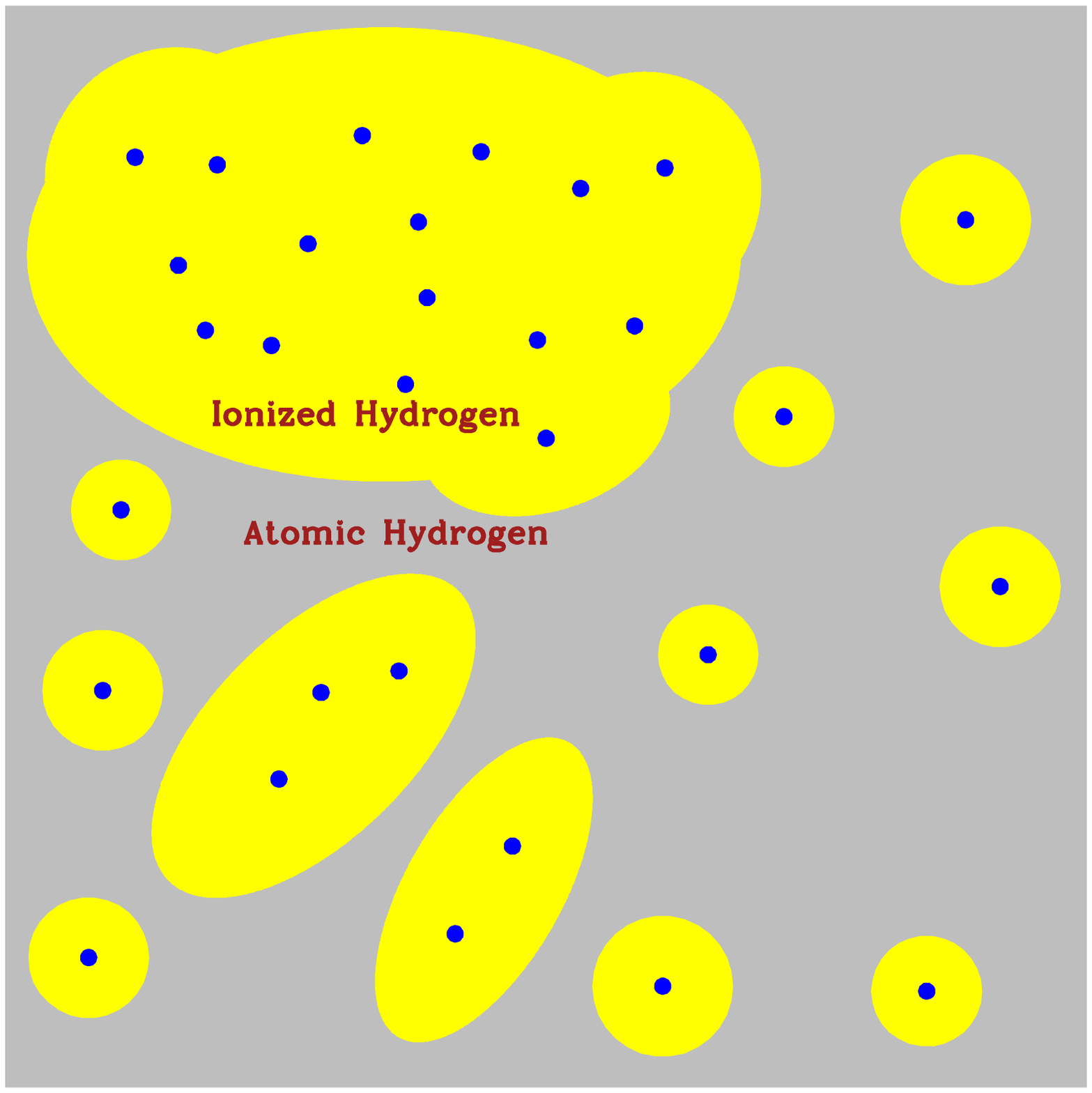}
\hfill
\includegraphics[width=2in]{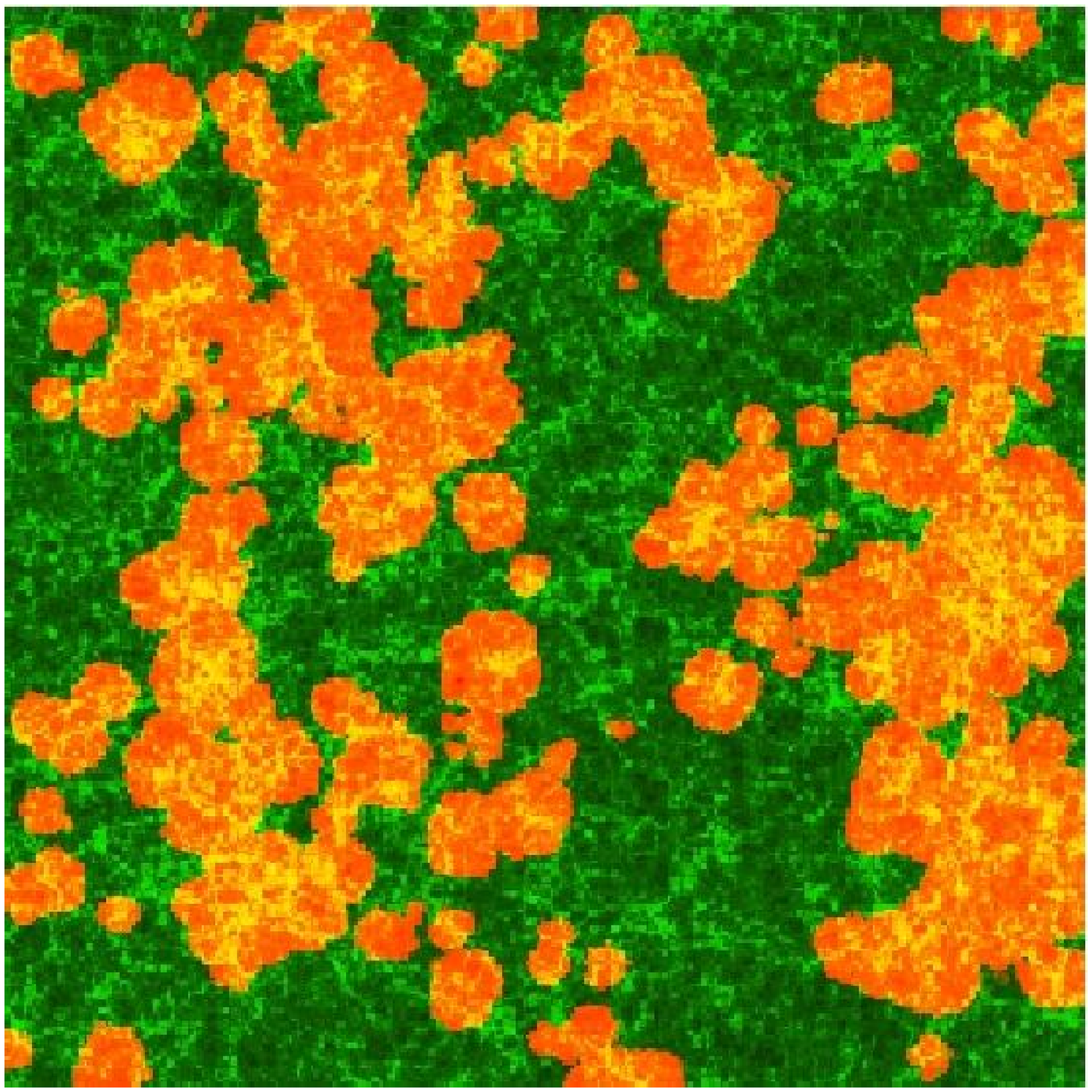}
\caption{The spatial structure of cosmic reionization. The illustration
(left panel, based on Barkana \& Loeb 2004b) shows how regions with
large-scale overdensities form large concentrations of galaxies (dots)
whose ionizing photons produce enormous joint ionized bubbles (upper
left). At the same time, galaxies are rare within large-scale voids, in
which the IGM is still mostly neutral (lower right). A numerical simulation
of reionization (right panel, from Mellema et al. 2006) indeed displays
such variation in the sizes of ionized bubbles (orange), shown overlayed on
the density distribution (green).}
\label{fig:rei}
\end{figure}

Several quasars beyond $z\sim6$ show in their spectra a Gunn-Peterson
trough, a blank spectral region at wavelengths shorter than Ly$\alpha$ at
the quasar redshift (Figure~\ref{fig:19qsos}). The detection of
Gunn-Peterson troughs indicates a rapid change \cite{Fan02,White03,Fan06}
in the neutral content of the IGM at $z\sim6$, and hence a rapid change in
the intensity of the background ionizing flux. However, even a small atomic
hydrogen fraction of $\sim 10^{-3}$ would still produce nearly complete
Ly$\alpha$ absorption.

A key point is that the spatial distribution of ionized bubbles is
determined by clustered groups of galaxies and not by individual
galaxies. At such early times galaxies were strongly clustered even on very
large scales (up to tens of Mpc), and these scales therefore dominated the
structure of reionization \cite{BL04b}. The basic idea is simple
\cite{Kaiser}. At high redshift, galactic halos are rare and correspond to
high density peaks. As an analogy, imagine searching on Earth for mountain
peaks above 5000 meters. The 200 such peaks are not at all distributed
uniformly but instead are found in a few distinct clusters on top of large
mountain ranges. Given the large-scale boost provided by a mountain range,
a small-scale crest need only provide a small additional rise in order to
become a 5000 meter peak. The same crest, if it formed within a valley,
would not come anywhere near 5000 meters in total height. Similarly, in
order to find the early galaxies, one must first locate a region with a
large-scale density enhancement, and then galaxies will be found there in
abundance.

The ionizing radiation emitted from the stars in each galaxy initially
produces an isolated ionized bubble. However, in a region dense with
galaxies the bubbles quickly overlap into one large bubble, completing
reionization in this region while the rest of the Universe is still mostly
neutral (Figure~\ref{fig:rei}). Most importantly, since the abundance of
rare density peaks is very sensitive to small changes in the density
threshold, even a large-scale region with a small enhanced density (say,
10\% above the mean density of the Universe) can have a much larger
concentration of galaxies than in other regions (e.g., a 50\%
enhancement). On the other hand, reionization is harder to achieve in dense
regions, since the protons and electrons collide and recombine more often
in such regions, and newly-formed hydrogen atoms need to be reionized again
by additional ionizing photons. However, the overdense regions end up
reionizing first since the number of ionizing sources in these regions is
increased so strongly \cite{BL04b,McQuinn07}. The large-scale topology of
reionization is therefore inside out, with underdense voids reionizing only
at the very end of reionization, with the help of extra ionizing photons
coming in from their surroundings (which have a higher density of galaxies
than the voids themselves). This is a key prediction awaiting observational
tests.

Detailed analytical models that account for large-scale variations in the
abundance of galaxies \cite{Fur04} confirm that the typical bubble size
starts well below a Mpc early in reionization, as expected for an
individual galaxy, rises to 5--10 comoving Mpc during the central phase
(i.e., when the Universe is half ionized), and then by another factor of
$\sim$5 towards the end of reionization. (These scales are given in
comoving units that scale with the expansion of the Universe, so that the
actual sizes at a redshift $z$ were smaller than these numbers by a factor
of $1+z$.) Numerical simulations have only recently begun to reach the
enormous scales needed to capture this evolution \cite{Ciardi,Mellema,
Zahn}. Accounting precisely for gravitational evolution over a wide range
of scales but still crudely for gas dynamics, star formation, and the
radiative transfer of ionizing photons, the simulations confirm that the
large-scale topology of reionization is inside out, and that this topology
can be used to study the abundance and clustering of the ionizing sources
(Figures~\ref{fig:rei} and \ref{fig:Mellema}).

Wyithe \& Loeb (2004b) \cite{WL04b} showed that the characteristic size of
the ionized bubbles at the end reionization can be calculated based on
simple considerations that only depend on the power-spectrum of density
fluctuations and the redshift. As the size of an ionized bubble increases,
the time it takes a 21-cm photon to traverse it gets longer. At the same
time, the variation in the time at which different regions reionize becomes
smaller as the regions grow larger. Thus, there is a maximum size above
which the photon crossing time is longer than the cosmic variance in
ionization time. Regions bigger than this size will be ionized at their
near side by the time a 21-cm photon will cross them towards the observer
from their far side. They would appear to the observer as one-sided, and
hence signal the end of reionization. These ``light cone'' considerations
imply a characteristic size for the ionized bubbles of $\sim 10$ physical
Mpc at $z\sim 6$ (equivalent to 70 comoving Mpc).  This result implies that
future radio experiments should be tuned to a characteristic angular scale
of tens of arcminutes and have a minimum frequency band-width of 5-10 MHz
for an optimal detection of 21-cm brightness fluctuations near the end of
reionization.

\subsection{Simulations of Reionization}

Simulating reionization is challenging for two reasons. First, one needs to
incorporate radiative transfer at multiple photon frequencies into a code
that follows the dynamics of gas and dark matter. This implies that the
sources of the radiation, i.e. galaxies, need to be resolved.  Second, one
needs to simulate a sufficiently large volume of the Universe for cosmic
variance not to play a role \cite{BL04b}. Towards the end of reionization,
the sizes of individual ionized regions grow up to a scale of $\sim
50$--100 comoving Mpc \cite{WL04b,Fur04} and the representative volume
needs to include many such region in order for it to fully describe the
large-scale topology of reionization. There is an obvious tension between
the above two requirements for simulating small scales as well as large
scales simultaneously.

Numerical simulations of reionization are starting to achieve the dynamic
range required to resolve galaxy halos across the scale of hundreds of
comoving Mpc, larger than the size of the ionized regions at the end of the
reionization process \cite{Zahn,Iliev,Trac}.  These simulations cannot
yet follow in detail the formation of individual stars within galaxies, or
the feedback that stars produce on the surrounding gas, such as
photoheating or the hydrodynamic and chemical impact of supernovae, which
blow hot bubbles of gas enriched with the chemical products of stellar
nucleosynthesis. Thus, the simulations cannot directly predict whether the
stars that form during reionization are similar to the stars in the Milky
Way and nearby galaxies or to the primordial $100 M_{\odot}$
behemoths. They also cannot determine whether feedback prevents low-mass
dark matter halos from forming stars. Thus, models are needed that make it
possible to vary all these astrophysical parameters of the ionizing
sources.

\section{Imaging Cosmic Hydrogen}

\subsection{Basic Principles}

\begin{figure}
\centering
\includegraphics[width=0.7\columnwidth,angle=-90]{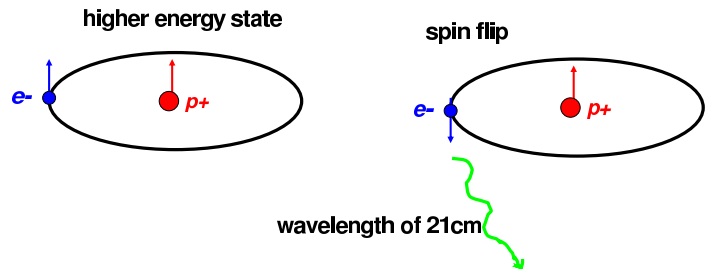}
\caption{The 21cm transition of hydrogen. The higher energy level the spin
of the electron (e-) is aligned with that of the proton (p+).  A spin flip
results in the emission of a photon with a wavelength of 21cm (or a
frequency of 1420MHz).}
\label{21cm}
\end{figure}

The ground state of hydrogen exhibits hyperfine splitting involving the
spins of the proton and the electron. The state with parallel spins (the
triplet state) has a slightly higher energy than the state with
anti-parallel spins (the singlet state). The 21-cm line associated with the
spin-flip transition from the triplet to the singlet state is often used to
detect neutral hydrogen in the local Universe. At high redshift, the
occurrence of a neutral pre-reionization intergalactic medium (IGM) offers
the prospect of detecting the first sources of radiation and probing the
reionization era by mapping the 21-cm absorption or emission from neutral
regions.  Regions where the gas is slightly denser than the mean would
produce a stronger signal. Therefore, the 21cm brightness will fluctuate
across the sky as a result of the inhomogeneous distribution of
hydrogen. Moreover, this resonant line can be used to slice the Universe at
different redshifts $z$ by observing different wavelengths corresponding to
$21{\rm cm} (1+z)$.  Altogether, the 21cm brightness fluctuation can be
used to map the inhomogeneous hydrogen distribution in three dimensions.

\begin{figure}
\centering
\includegraphics[height=6cm]{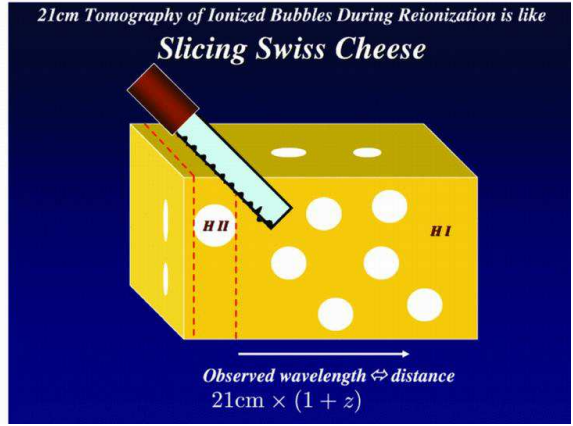}
\caption{21cm imaging of ionized bubbles during the epoch of reionization
is analogous to slicing swiss cheese. The technique of slicing at intervals
separated by the typical dimension of a bubble is optimal for revealing
different pattens in each slice.}  \label{swiss}
\end{figure}

The atomic hydrogen gas formed soon after the big-bang, was affected by
processes ranging from quantum fluctuations during the early epoch of
inflation to irradiation by the first galaxies at late times. Mapping this
gas through its resonant 21cm line serves a dual role as a powerful probe
of fundamental physics and of astrophysics. The facets of fundamental
physics include the initial density fluctuations imprinted by inflation as
well as the nature of the dark matter, which amplifies these fluctuations
during the matter-dominated era. It is possible to avoid the contamination
from astrophysical sources by observing the Universe before the first
galaxies had formed.  In the concordance $\Lambda$CDM cosmological model,
the 21cm brightness fluctuations of hydrogen were shaped by fundamental
physics (inflation, dark matter, and atomic physics) at redshifts $z> 20$,
and by the radiation from galaxies at lower redshifts.

Following cosmological recombination at $z\sim 10^3$, the residual fraction
of free electrons coupled the gas thermally to the cosmic microwave
background (CMB) for another 65 million years ($z\sim 200$), but afterwards
the gas decoupled and cooled faster than the CMB through its cosmic
expansion.  In the redshift interval of the so-called {\it Dark Ages}
before the first stars had formed, $30< z< 200$, the {\it spin
temperature} of hydrogen, $T_s$ (defined through the level population of
the spin-flip transition), was lower than the CMB temperature, $T_\gamma$,
and the gas appeared in absorption.  The primordial inhomogeneities of the
gas produced varying levels of 21cm absorption and hence brightness
fluctuations. Detection of this signal can be used to constrain models of
inflation as well as the nature of dark matter \cite{Hogan,Scott,LZ04,L07}.
Altogether, there are $\sim 10^{16}$ independent pixels on the 21cm sky
from this epoch (instead of $\sim 10^7$ for the CMB). They make the richest
data set on the sky, providing an unprecedented probe of non-Gaussianity
and running of the spectral index of the power-spectrum of primordial
density fluctuations from inflation. Detection of these small-scale
fluctuations can also be used to infer the existence of massive neutrinos
and other sub-dominant components in addition to the commonly inferred cold
dark matter particles.

\begin{figure}
\centering
\includegraphics[height=10cm,angle=-90]{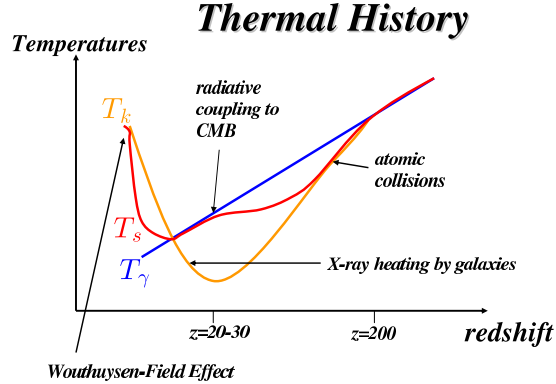}
\caption{Schematic sketch of the evolution of the kinetic temperature
($T_k$) and spin temperature ($T_s$) of cosmic hydrogen (from Loeb
2006). Following cosmological recombination at $z\sim 10^3$, the gas
temperature (orange curve) tracks the CMB temperature (blue line;
$T_\gamma\propto (1+z)$) down to $z\sim 200$ and then declines below it
($T_k\propto (1+z)^2$) until the first X-ray sources (accreting black holes
or exploding supernovae) heat it up well above the CMB temperature. The
spin temperature of the 21cm transition (red curve) interpolates between
the gas and CMB temperatures. Initially it tracks the gas temperature
through collisional coupling; then it tracks the CMB through radiative
coupling; and eventually it tracks the gas temperature once again after the
production of a cosmic background of UV photons between the Ly$\alpha$ and
the Lyman-limit frequencies that redshift or cascade into the Ly$\alpha$
resonance (through the Wouthuysen-Field effect \cite{Wou,Field}). Parts of
the curve are exaggerated for pedagogical purposes.  The exact shape depends
on astrophysical details about the first galaxies, such as the production
of X-ray binaries, supernovae, nuclear accreting black holes, and the
generation of relativistic electrons in collisionless shocks which produce
UV and X-ray photons through inverse-Compton scattering of CMB photons. }
\label{spin}
\end{figure}

After the first galaxies formed and X-ray sources heated the gas above the
CMB temperature, the gas appeared in 21cm emission. The bubbles of ionized
hydrogen around groups of galaxies were dark and dominated the 21cm
fluctuations \cite{FOB,PF}. After the first stars had formed, the 21cm
fluctuations were sourced mainly by the hydrogen ionized fraction, and spin
temperature \cite{Madau97}.
 
The basic physics of the hydrogen spin transition is determined as follows
(for a more detailed treatment, see Refs. \cite{Madau97,FOB}). The
ground-state hyperfine levels of hydrogen tend to thermalize with the CMB
bringing the IGM away from thermal equilibrium, then the gas becomes
observable against the CMB in emission or in absorption. The relative
occupancy of the spin levels is described in terms of the hydrogen spin
temperature $T_S$, defined through the Boltzman factor, 
\begin{equation}
\frac{n_1}{n_0}=3\, \exp\left\{-\frac{T_*}{T_S}\right\}\ , 
\end{equation} 
where $n_0$ and $n_1$ refer respectively to the singlet and triplet
hyperfine levels in the atomic ground state ($n=1$), and $T_*=0.068$ K is
defined by $k_B T_*=E_{21}$, where the energy of the 21 cm transition is
$E_{21}=5.9 \times 10^{-6}$ eV, corresponding to a frequency of 1420
MHz. In the presence of the CMB alone, the spin states reach thermal
equilibrium with the CMB temperature $T_S=T_{\gamma}=2.725 (1+z)$ K on a
time-scale of $T_*/(T_{\gamma} A_{10}) \simeq 3 \times 10^5 (1+z)^{-1}$
yr, where $A_{10}=2.87 \times 10^{-15}$ s$^{-1}$ is the spontaneous decay
rate of the hyperfine transition. This time-scale is much shorter than the
age of the Universe at all redshifts after cosmological recombination.

The IGM is observable when the kinetic temperature $T_k$ of the gas differs
from the CMB temperature $T_{\gamma}$ and an effective mechanism couples
$T_S$ to $T_k$. Collisional de-excitation of the triplet level
\cite{Purcell} dominates at very high redshift, when the gas density (and
thus the collision rate) is still high, making the gas observable in
absorption.  Once a significant galaxy population forms in the Universe,
the X-rays they emit heat $T_k$ above $T_{\gamma}$ and the UV photons they
emit couple $T_s$ to $T_k$ making the gas appear in 21cm emission.  The
latter coupling mechanism acts through the scattering of Ly$\alpha$ photons
\cite{Wou,Field}. Continuum UV photons produced by early radiation sources
redshift by the Hubble expansion into the local Ly$\alpha$ line at a lower
redshift. These photons mix the spin states via the Wouthuysen-Field
process whereby an atom initially in the $n=1$ state absorbs a Ly$\alpha$
photon, and the spontaneous decay which returns it from $n=2$ to $n=1$ can
result in a final spin state which is different from the initial one. Since
the neutral IGM is highly opaque to resonant scattering, and the Ly$\alpha$
photons receive Doppler kicks in each scattering, the shape of the
radiation spectrum near Ly$\alpha$ is determined by $T_k$ \cite{Field}, and
the resulting spin temperature (assuming $T_S \gg T_*$) is then a weighted
average of $T_k$ and $T_{\gamma}$:
\begin{equation} 
T_S=\frac{T_{\gamma} T_k (1+x_{\rm tot}) }{T_k + T_{\gamma}
x_{\rm tot}}\ , 
\end{equation} 
where $x_{\rm tot} = x_{\alpha} + x_c$ is the sum of the radiative and
collisional threshold parameters. These parameters are $x_{\alpha} =
{{P_{10} T_\star}\over {A_{10} T_{\gamma}}}$, and $x_c = {{4
\kappa_{1-0}(T_k)\, n_H T_\star}\over {3 A_{10} T_{\gamma}}}$, where
$P_{10}$ is the indirect de-excitation rate of the triplet $n=1$ state via
the Wouthuysen-Field process, related to the total scattering rate
$P_{\alpha}$ of Ly$\alpha$ photons by $P_{10}=4 P_{\alpha}/27$
\cite{Field58}. Also, the atomic coefficient $\kappa_{1-0}(T_k)$ is
tabulated as a function of $T_k$ \cite{Allison,Zygelman}. Note that we have
adopted the modified notation (i.e., in terms of $x_\alpha$ and $x_c$) of
Barkana \& Loeb (2005b) \cite{BL05b}. The coupling of the spin temperature
to the gas temperature becomes substantial when $x_{\rm tot} > 1$; in
particular, $x_{\alpha} = 1$ defines the thermalization rate \cite{Madau97}
of $P_{\alpha}$: $P_{\rm th} \equiv \frac{27 A_{10} T_{\gamma}}{4 T_*}
\simeq 7.6 \times 10^{-12}\, \left(\frac{1+z}{10}\right)\ {\rm s}^{-1}$.

A patch of neutral hydrogen at the mean density and with a uniform
$T_S$ produces (after correcting for stimulated emission) an optical
depth at a present-day (observed) wavelength of $21 (1+z)$ cm,
\begin{equation} 
\tau(z) = 9.0
\times 10^{-3} \left(\frac{T_{\gamma}} {T_S} \right) \left (
\frac{\Omega_b h} {0.03} \right) \left(\frac{\Omega_m}{0.3}\right)^ {-1/2}
\left(\frac{1+z}{10}\right)^{1/2}\ , 
\end{equation} 
assuming a high redshift $z\gg1$. The observed spectral intensity $I_{\nu}$
relative to the CMB at a frequency $\nu$ is measured by radio astronomers
as an effective brightness temperature $T_b$ of blackbody emission at this
frequency, defined using the Rayleigh-Jeans limit of the Planck radiation
formula: $I_{\nu} \equiv 2 k_B T_b \nu^2 / c^2 $.

The brightness temperature through the IGM is $T_b=T_{\gamma}
e^{-\tau}+T_S (1-e^{-\tau})$, so the observed differential antenna
temperature of this region relative to the CMB is \cite{Madau97}
\begin{eqnarray} 
T_b&=&(1+z)^{-1} (T_S-T_{\gamma}) (1-e^{-\tau}) \nonumber \\ &\simeq& 28\,
{\rm mK}\, \left( \frac{\Omega_b h} {0.033} \right)
\left(\frac{\Omega_m}{0.27}\right)^ {-1/2} \left( \frac{1+z} {10}
\right)^{1/2} \left( \frac{T_S-T_{\gamma}} {T_S} \right)\ ,
\end{eqnarray} 
where $\tau \ll 1$ can be assumed and $T_b$ has been redshifted to redshift
zero. Note that the combination that appears in $T_b$ is ${T_S - T_{\gamma}
\over T_S} = {x_{\rm tot}\over 1+ x_{\rm tot}} \left(1 - {T_{\gamma}\over
T_k} \right)$.  In overdense regions, the observed $T_b$ is proportional to
the overdensity, and in partially ionized regions $T_b$ is proportional to
the neutral fraction. Also, if $T_S \gg T_{\gamma}$ then the IGM is
observed in emission at a level that is independent of $T_S$. On the other
hand, if $T_S \ll T_{\gamma}$ then the IGM is observed in absorption at a
level that is enhanced by a factor of $T_{\gamma} / T_S$.

To complement computationally-intensive simulations of reionization,
various groups developed approximate schemes for simulating 21-cm maps in
the regime where $T_S$ is much larger than $T_\gamma$. For example,
Furlanetto et al. (2004) \cite{Fur04} developed an analytical model that
allows the calculation of the probability distribution (at a given
redshift) of the size of the ionizing bubble surrounding a random point in
space. Zahn et al. (2006) \cite{Zahn} have considered numerical schemes
that apply the Furlanetto et al. (2004) model to either the initial
conditions of their simulation or to part of its results
(Figure~\ref{fig:Zahn}).

\begin{figure}
\centering
\includegraphics[height=7cm]{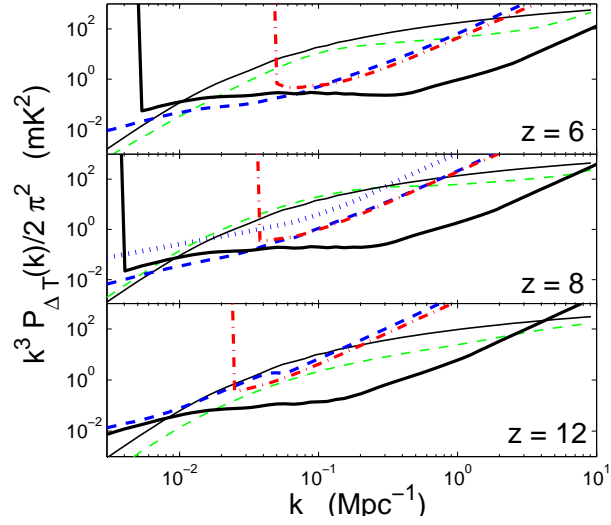}
\caption{Detectability of the power-spectrum of 21cm fluctuations by
different future observatories (from McQuinn et al. 2006).  The detector
noise plus sample variance errors is shown for a $1000$ hr observation on a
single field in the sky, assuming perfect foreground removal, for MWA ({\it
thick dashed curve}), LOFAR ({\it thick dot-dashed curve}), and SKA ({\it
thick solid curve}) for wavenumber bins of $\Delta k = 0.5 \, k$.  The thin
solid curve represents the spherically averaged signal for a small
ionization fraction and $T_s \gg T_{\gamma}$.  }
\label{fig:errors}
\end{figure}

\subsection{Predicted 21cm Signal}

\begin{figure}
\centering
\includegraphics[width=4.6in]{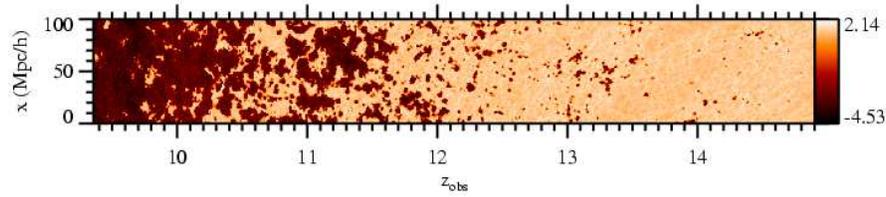}
\caption{Close-up of cosmic evolution during the epoch of reionization, as
revealed in a predicted 21-cm map of the IGM based on a numerical
simulation (from Mellema et al. 2006). This map is constructed from slices
of the simulated cubic box of side 150 Mpc (in comoving units), taken at
various times during reionization, which for the parameters of this
particular simulation spans a period of 250 million years from redshift 15
down to 9.3. The vertical axis shows position $\chi$ in units of Mpc/h
(where the Hubble constant in units of $100~{\rm km~s^{-1}}$ is
$h=0.7$). This two-dimensional slice of the sky (one linear direction on
the sky versus the line-of-sight or redshift direction) shows
$\log_{10}(T_b)$, where $T_b$ (in mK) is the 21-cm brightness temperature
relative to the CMB. Since neutral regions correspond to strong emission
(i.e., a high $T_b$), this slice illustrates the global progress of
reionization and the substantial large-scale spatial fluctuations in
reionization history. Observationally it corresponds to a narrow strip half
a degree in length on the sky observed with radio telescopes over a
wavelength range of 2.2 to 3.4 m (with each wavelength corresponding to
21-cm emission at a specific line-of-sight distance and redshift).}
\label{fig:Mellema}
\end{figure}

\begin{figure}
\centering
\includegraphics[width=4.6in]{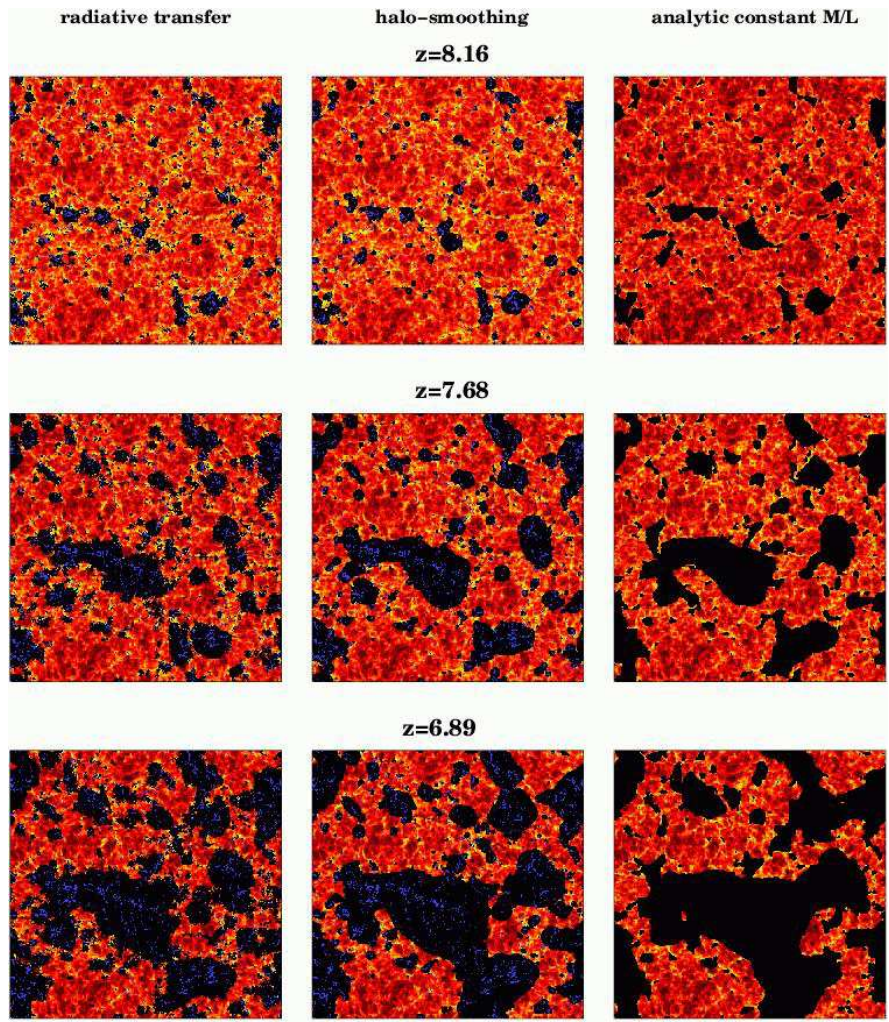}
\caption{Maps of the 21-cm brightness temperature comparing results of a
numerical simulation and of two simpler numerical schemes, at three
different redshifts (from Zahn et al. 2006). Each map is 65.6 Mpc/$h$ on a
side, with a depth (0.25 Mpc/$h$) that is comparable to the frequency
resolution of planned experiments. The ionized fractions are $x_{\rm
i}=0.13$, 0.35 and 0.55 for $z=8.16$, 7.26 and 6.89, respectively. All
three maps show a very similar large-scale ionization topology. \emph{Left
column:} Numerical simulation, showing the ionized bubbles (black) produced
by the ionizing sources (blue dots) that form in the
simulation. \emph{Middle column:} Numerical scheme that applies the
Furlanetto et al. (2004) analytical model to the final distribution of
ionizing sources that form in the simulation. \emph{Right column:}
Numerical scheme that applies the Furlanetto et al. (2004) analytical model
to the linear density fluctuations that are the initial conditions of the
simulation.}
\label{fig:Zahn}
\end{figure}

In approaching redshifted 21-cm observations, although the first inkling
might be to consider the mean emission signal, the signal is orders of
magnitude fainter than foreground synchrotron emission from relativistic
electrons in the magnetic field of our own Milky Way \cite{FOB} as well as
other galaxies \cite{Di}. Thus cosmologists have focused on the expected
characteristic variations in $T_b$, both with position on the sky and
especially with frequency, which signifies redshift for the cosmic
signal. The synchrotron foreground is expected to have a smooth frequency
spectrum, and so it is possible to isolate the cosmological signal by
taking the difference in the sky brightness fluctuations at slightly
different frequencies (as long as the frequency separation corresponds to
the characteristic size of ionized bubbles). The 21-cm brightness
temperature depends on the density of neutral hydrogen. As explained in the
previous subsection, large-scale patterns in the reionization are driven by
spatial variations in the abundance of galaxies; the 21-cm fluctuations
reach $\sim$5 mK (root mean square) in brightness temperature
(Figure~\ref{fig:Mellema}) on a scale of 10 comoving Mpc. While detailed
maps will be difficult to extract due to the foreground emission, a
statistical detection of these fluctuations (through the power spectrum) is
expected to be well within the capabilities of the first-generation
experiments now being built \cite{Bowman,McQuinn}.  Current work suggests
that the key information on the topology and timing of reionization can be
extracted statistically (Fig. \ref{fig:errors}).

The theoretical expectations for reionization and for the 21-cm signal are
based on rather large extrapolations from observed galaxies to deduce the
properties of much smaller galaxies that formed at an earlier cosmic
epoch. Considerable surprises are thus possible, such as an early
population of quasars or even unstable exotic particles that emitted
ionizing radiation as they decayed. 


An important cross-check on these measurements is possible by measuring the
particular form of anisotropy, expected in the 21-cm fluctuations, that is
caused by gas motions along the line of sight
\cite{Kaiser87,Bharadwaj,BL05a}. This anisotropy, expected in any
measurement of density that is based on a spectral resonance or on redshift
measurements, results from velocity compression. Consider a photon
traveling along the line of sight that resonates with absorbing atoms at a
particular point. In a uniform, expanding Universe, the absorption optical
depth encountered by this photon probes only a narrow strip of atoms, since
the expansion of the Universe makes all other atoms move with a relative
velocity that takes them outside the narrow frequency width of the
resonance line. If there is a density peak, however, near the resonating
position, the increased gravity will reduce the expansion velocities around
this point and bring more gas into the resonating velocity width. This
effect is sensitive only to the line-of-sight component of the velocity
gradient of the gas, and thus causes an observed anisotropy in the power
spectrum even when all physical causes of the fluctuations are
statistically isotropic. Barkana \& Loeb (2005a) showed that this
anisotropy is particularly important in the case of 21-cm
fluctuations. When all fluctuations are linear, the 21-cm power spectrum
takes the form \cite{BL05a} 
$P_{\rm 21-cm}({\bf k}) = \mu^4 P_{\rho}(k) + 2
\mu^2 P_{\rho - {\rm iso}} (k) + P_{\rm iso}$ , where $\mu = \cos\theta$ 
in terms of the angle $\theta$ between the wavevector ${\bf k}$ of
a given Fourier mode and the line of sight, $P_{\rm iso}$ is the isotropic
power spectrum that would result from all sources of 21-cm fluctuations
without velocity compression, $P_{\rho}(k)$ is the 21-cm power spectrum
from gas density fluctuations alone, and $P_{\rho - {\rm iso}} (k)$ is the
Fourier transform of the cross-correlation between the density and all
sources of 21-cm fluctuations. The three power spectra can also be denoted
$P_{\mu^4}(k)$, $P_{\mu^2}(k)$, and $P_{\mu^0}(k)$, according to the power
of $\mu$ that multiplies each term. At these redshifts, the 21-cm
fluctuations probe the infall of the baryons into the dark matter potential
wells \cite{BL05c}. The power spectrum shows remnants of the photon-baryon
acoustic oscillations on large scales, and of the baryon pressure
suppression on small scales \cite{Naoz}.

Once stellar radiation becomes significant, many processes can contribute
to the 21-cm fluctuations. The contributions include fluctuations in gas
density, temperature, ionized fraction, and Ly$\alpha$ flux. These
processes can be divided into two broad categories: The first, related to
{\it ``physics''}, consists of probes of fundamental, precision cosmology,
and the second, related to {\it ``astrophysics''}, consists of probes of
stars. Both categories are interesting -- the first for precision measures
of cosmological parameters and studies of processes in the early Universe,
and the second for studies of the properties of the first
galaxies. However, the astrophysics depends on complex non-linear processes
(collapse of dark matter halos, star formation, supernova feedback), and
must be cleanly separated from the physics contribution, in order to allow
precision measurements of the latter.  As long as all the fluctuations are
linear, the anisotropy noted above allows precisely this separation of the
physics from the astrophysics of the 21-cm fluctuations \cite{BL05a}. In
particular, the $P_{\mu^4}(k)$ is independent of the effects of stellar
radiation, and is a clean probe of the gas density fluctuations. Once
non-linear terms become important, there arises a significant mixing of the
different terms; in particular, this occurs on the scale of the ionizing
bubbles during reionization \cite{McQuinn}.

At early times, the 21-cm fluctuations are also affected by fluctuations in
the Ly$\alpha$ flux from stars, a result that yields an indirect method to
detect and study the early population of galaxies at $z \sim 20$
\cite{BL05b}. The fluctuations are caused by biased inhomogeneities in the
density of galaxies, along with Poisson fluctuations in the number of
galaxies. Observing the power-spectra of these two sources would probe the
number density of the earliest galaxies and the typical mass of their host
dark matter halos. Furthermore, the enhanced amplitude of the 21cm
fluctuations from the era of Ly$\alpha$ coupling improves considerably the
practical prospects for their detection. Precise predictions account for
the detailed properties of all possible cascades of a hydrogen atom after
it absorbs a photon \cite{Hirata,PFa}. Around the same time, X-rays may
also start to heat the cosmic gas, producing strong 21-cm fluctuations due
to fluctuations in the X-ray flux \cite{PF}.

\subsection{Baryonic Acoustic Oscillation}

\begin{figure}
\includegraphics[width=4.7in]{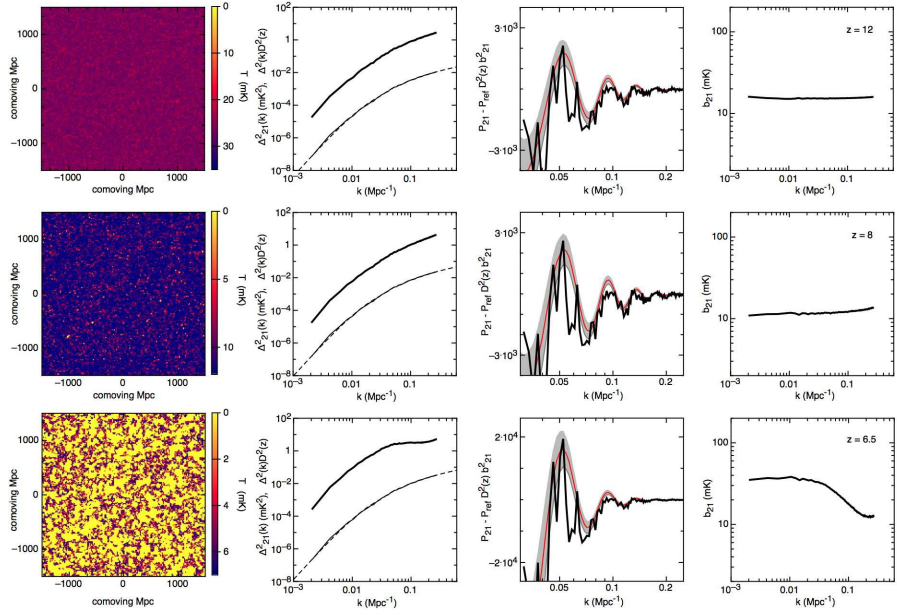}
\caption{Examples of the 21cm power spectra during reionization (from
Wyithe, Loeb, \& Geil 2007). {\em Left panels:} Maps of the 21cm emission
from slices through the numerical simulation boxes, each 3000 co-moving Mpc
on a side with a thickness of 12 co-moving Mpc. In these maps yellow
designates the absence of redshifted 21cm emission. {\em Central-left
panels:} The corresponding matter power spectra multiplied by the growth
factor squared (thin solid lines) as well as the 21cm (thick solid lines)
power spectra computed from the simulation box. The input co-moving power
spectrum $P$ (also multiplied by the growth factor squared) is shown for
comparison (short-dashed lines). {\em Central-right panels:} The baryonic
oscillations component of the simulated 21cm power spectrum. The curves
(thick dark lines) show the difference between the simulated 21cm power
spectrum, and a reference no wiggle 21cm power spectrum computed from the
theoretical no wiggle reference matter power spectrum multiplied by the
square of the product between the bias and the growth factor
[i.e. $P_{21}-P_{\rm ref}b_{21}^2D^2$].  For comparison, the red lines show
the difference between the input matter and the no-BAO reference matter
power spectra, multiplied by the bias and growth factor squared
[i.e. $(P-P_{\rm ref})b_{21}^2D^2$]. The grey band surrounding this curve
shows the level of statistical scatter in realizations of the power
spectrum due to the finite size of the simulation volume. {\em Right
panels:} The scale dependent bias ($b_{21}$).  The upper, central and lower
panels show results at $z=12$, $z=8$ and $z=6.5$, which have global neutral
fractions of 98\%, 48\% and 11\% respectively in the model shown.  }
\label{fig2}
\end{figure}

As described above, the fluctuations in the emission of redshifted 21cm
photons from neutral intergalactic hydrogen will provide an unprecedented
probe of the reionization era.  Conventional wisdom assumes that this 21cm
signal disappears as soon as reionization is complete, when little atomic
hydrogen is left through most of the volume of the IGM.  However, the
statistics of damped Ly$\alpha$ absorbers (DLAs) in quasar spectra indicate
that a few percent of the baryonic mass reservoir remains in the form of
atomic hydrogen at all redshifts.  The residual hydrogen is self-shielded
from UV radiation within dense regions in which the recombination rate is
high. Wyithe \& Loeb (2007) used a physically-motivated model to show that
residual neutral gas would generate a significant post-reionization 21cm
signal. Even though the signal is much weaker (see Fig. \ref{fig2}) than
that of a fully-neutral IGM, the synchrotron foreground, whose brightness
temperature scales as $\propto (1+z)^{2.6}$, is also much weaker at the
corresponding low redshifts.  Thus, the power-spectrum of fluctuations in
this signal will be detectable by the first generation of low-frequency
observatories at a signal-to-noise that is comparable to that achievable in
observations of the reionization era.  The statistics of 21cm fluctuations
will therefore probe not only the pre-reionization IGM, but rather the
entire process of HII region overlap, as well as the appearance of the
diffuse ionized IGM. With an angular resolution of an arcminute, the radio
beam of future interferometers will contain many DLAs and will not resolve
them individually. Rather, the observations will map the course-binned
distribution of neutral hydrogen across scales of tens of comoving Mpc.

Wyithe, Loeb, \& Geil (2007) demonstrated that the power spectrum of the
cumulative 21cm emission during and after reionization will show baryonic
acoustic oscillations (BAOs), whose comoving scale can be used as a
standard ruler to infer the evolution of the equation of state for the dark
energy.  The BAO yardstick can be used to measure the dependence of both
the angular diameter distance and Hubble parameter on redshift. The
wavelength of the BAO is related to the size of the sound horizon at
recombination, as this reflects the distance out to which different points
were correlated in the radiation-baryon fluid. Its value depends on the
Hubble constant and on the matter and baryon densities. However, it does
not depend on the amount or nature of the dark energy. Thus, measurements
of the angular diameter distance and Hubble parameter can in turn be used
to constrain the possible evolution of the dark energy with cosmic
time. This idea was originally proposed in relation to galaxy redshift
surveys \cite{Blake,Hu,Seo} and has since received significant theoretical
attention \cite{Glazebrook,Seo05,Angulo}. Moreover, measurement of the BAO
scale has been achieved within large surveys of galaxies at low redshift,
illustrating its potential \cite{Cole,Eisenstein}.  Galaxy redshift surveys
are best suited to studies of the dark energy at relatively late times due
to the difficulty of obtaining accurate redshifts for large numbers of high
redshift galaxies. Wyithe, Loeb, \& Geil (2007) have found that the first
generation of low-frequency experiments (such as MWA or LOFAR) will be able
to constrain the acoustic scale to within a few percent in a redshift
window just prior to the end of the reionization era. This sensitivity to
the acoustic scale is comparable to the best current measurements from
galaxy redshift surveys, but at much higher redshifts. Future extensions of
the first generation experiments (involving an order of magnitude increase
in the antennae number of the MWA) could reach sensitivities below one
percent in several redshift windows and could be used to study the dark
energy in the unexplored redshift regime of $3.5< z< 12$. Moreover, new
experiments with antennae designed to operate at higher frequencies would
allow precision measurements ($<1\%$) of the acoustic peak to be made at
more moderate redshifts ($1.5< z<3.5$), where they would be competitive
with ambitious spectroscopic galaxy surveys covering more than 1000 square
degrees. Together with other data sets, observations of 21cm fluctuations
will allow full coverage of the acoustic scale from the present time out to
$z\sim 12$ \cite{WL07b} and beyond \cite{BL05c}.

The left hand panels of Figure~\ref{fig2} show the 21cm emission from 12Mpc
slices through a numerical simulation \cite{WL07b}.  The higher redshift
example ($z=12$) is early in the reionization era, and shows no HII regions
forming at the resolution of the simulation (i.e. the IGM does not contain
ionized bubbles with radii $>5$ co-moving Mpc). The fluctuations in the
21cm emission are dominated by the density field at this time. The central
redshift ($z\sim8$) shows the IGM midway through the reionization process,
and includes a few HII regions above the simulation resolution. The lower
redshift example is just prior to the overlap of the ionized regions (and
hence the completion of reionization), when the IGM is dominated by large
percolating HII regions.

\subsection{Low-Frequency Arrays}

\begin{figure}
\centering
\includegraphics[height=7cm]{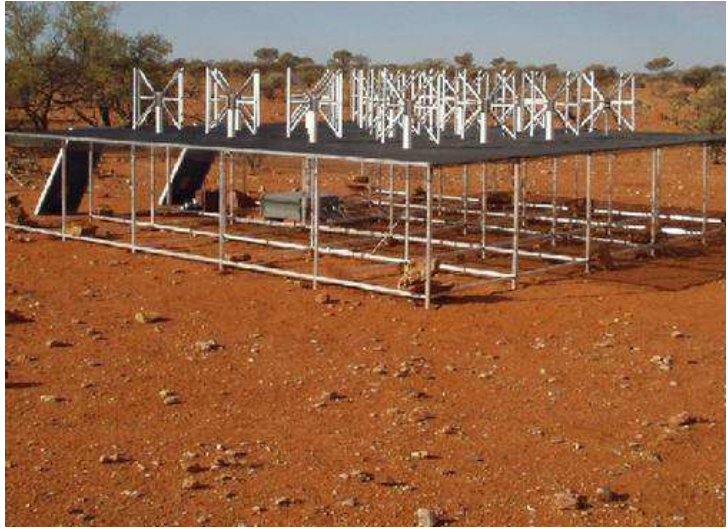}
\caption{Prototype of the tile design for the {\it Murchison Wide-Field
Array} (MWA) in western Australia, aimed at detecting redshifted 21cm from
the epoch of reionization. Each 4m$\times$4m tile contains 16 dipole
antennas operating in the frequency range of 80--300MHz. Altogether the
initial phase of MWA (the so-called ``Low-Frequency Demonstrator'') will
include 500 antenna tiles with a total collecting area of 8000 m$^2$ at
150MHz, scattered across a 1.5 km region and providing an angular
resolution of a few arcminutes.}
\label{MWA}
\end{figure}

The main obstacle towards detecting the 21cm signal is the synchrotron
foreground contamination from our Galaxy and extragalactic point sources,
whose brightness temperature rises steeply towards lower frequencies
($\propto \nu^{-2.6}$) and makes the detection of the redshifted 21cm line
more challenging at higher redshifts. This fact directed most experimental
and theoretical work so far towards the study of the astrophysics-dominated
era at low redshifts ($z< 20$), during which the 21cm fluctuations were
sourced mainly by the growth of ionized bubbles around galaxies. For
example, the Murchison Wide-Field Array (MWA; Fig. \ref{MWA}), which is
currently funded by NSF and the Australian government, is designed to cover
the redshift range of 6--17. The first generation MWA-demonstrator will
have 512 antenna tiles of $4\times 4$ dipole antennae each.

The prospect of studying reionization by mapping the distribution of atomic
hydrogen across the Universe using its prominent 21-cm spectral line has
motivated several teams to design and construct arrays of low-frequency
radio telescopes; the Low Frequency Array\footnote{http://www.lofar.org/},
MWA\footnote{\it
http://www.haystack.mit.edu/ast/arrays/mwa/index.html}, the
21CMA\footnote{\it http://arxiv.org/abs/astro-ph/0502029}, and ultimately
the Square Kilometer Array\footnote{\it http://www.skatelescope.org} will
search over the next decade for 21-cm emission or absorption from $z\sim
3.5$--15, redshifted and observed today at relatively low frequencies which
correspond to wavelengths of 1 to 4 meters.  Producing resolved images even
of large sources such as cosmological ionized bubbles requires telescopes
which have a kilometer scale. It is much more cost-effective to use a large
array of thousands of simple antennas distributed over several kilometers,
and to use computers to cross-correlate the measurements of the individual
antennas and combine them effectively into a single large telescope. The
new experiments are being placed mostly in remote sites, because the cosmic
wavelength region overlaps with more mundane terrestrial
telecommunications.

\section{Imaging Galaxies}

\subsection{Future Infrared Telescopes}

Narrow-band searches for redshifted Ly$\alpha$ emission have discovered
galaxies robustly out to redshifts $z=6.96$ \cite{Iye06} and potentially
out to $z=10$ \cite{Stark}. Existing observations provide a first glimpse
into the formation of the first galaxies \cite{MR02,Ka06,Taniguchi,DW07}
with potential theoretical implications for the epoch of reionization
\cite{Loeb99,RL99,Haiman99,MR04,SLE,McQuinn07, Mesinger07}.Future surveys
intend to exploit this search strategy further by pushing to even higher
redshifts and fainter flux levels
\cite{Horton04,Willis06,Willis07,Cuby07}. The spectral break due to
Ly$\alpha$ absorption by the IGM allows to identify high-redshift galaxies
photometrically with even greater sensitivity \cite{B1,B2,B3} (see overview
in Ref. \cite{Ellis} and phenomenological model in Ref. \cite{SLE}).

The construction of large infrared telescopes on the ground and in space
will provide us with new photos of first generation of galaxies during the
epoch of reionization.  Current plans include the space telescope JWST
(which will not be affected by the atmospheric background) as well as
ground-based telescopes which are 24-42 meter in diameter (such as the
GMT\footnote{http://www.gmto.org/}, TMT\footnote{http://celt.ucolick.org/},
and EELT\footnote{http://www.eso.org/projects/e-elt/}.  

\begin{figure}
\centering
\includegraphics[width=4in]{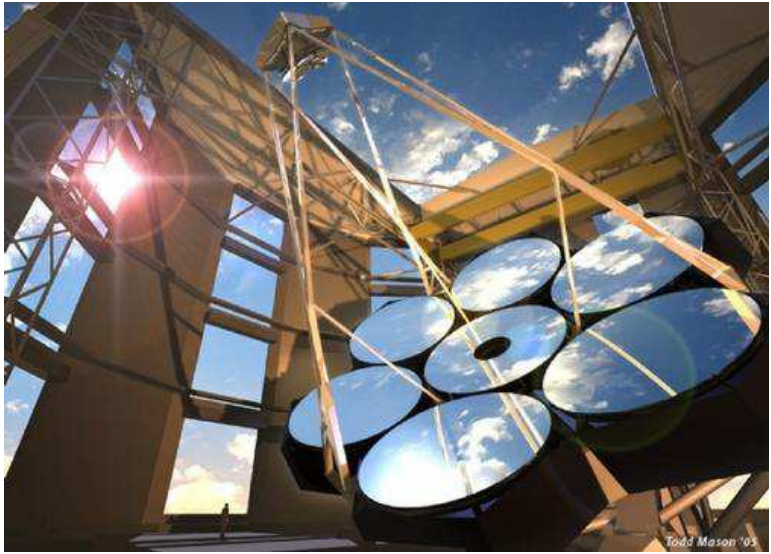}
\caption{Artist's conception of the design for one of the future giant
telescopes that could probe the first generation of galaxies from the
ground. The {\it Giant Magellan Telescope}\/ ({\it GMT}\/) will
contain seven mirrors (each 8.4 meter in diameter) and will have a
resolving power equivalent to a 24.5 meter (80 foot) primary
mirror. For more details see http://www.gmto.org/ .}
\label{gmt}
\end{figure}

\subsection{Cross-Correlating Galaxies with 21cm Maps}

\begin{figure}
\centering 
\includegraphics[width=4.6in]{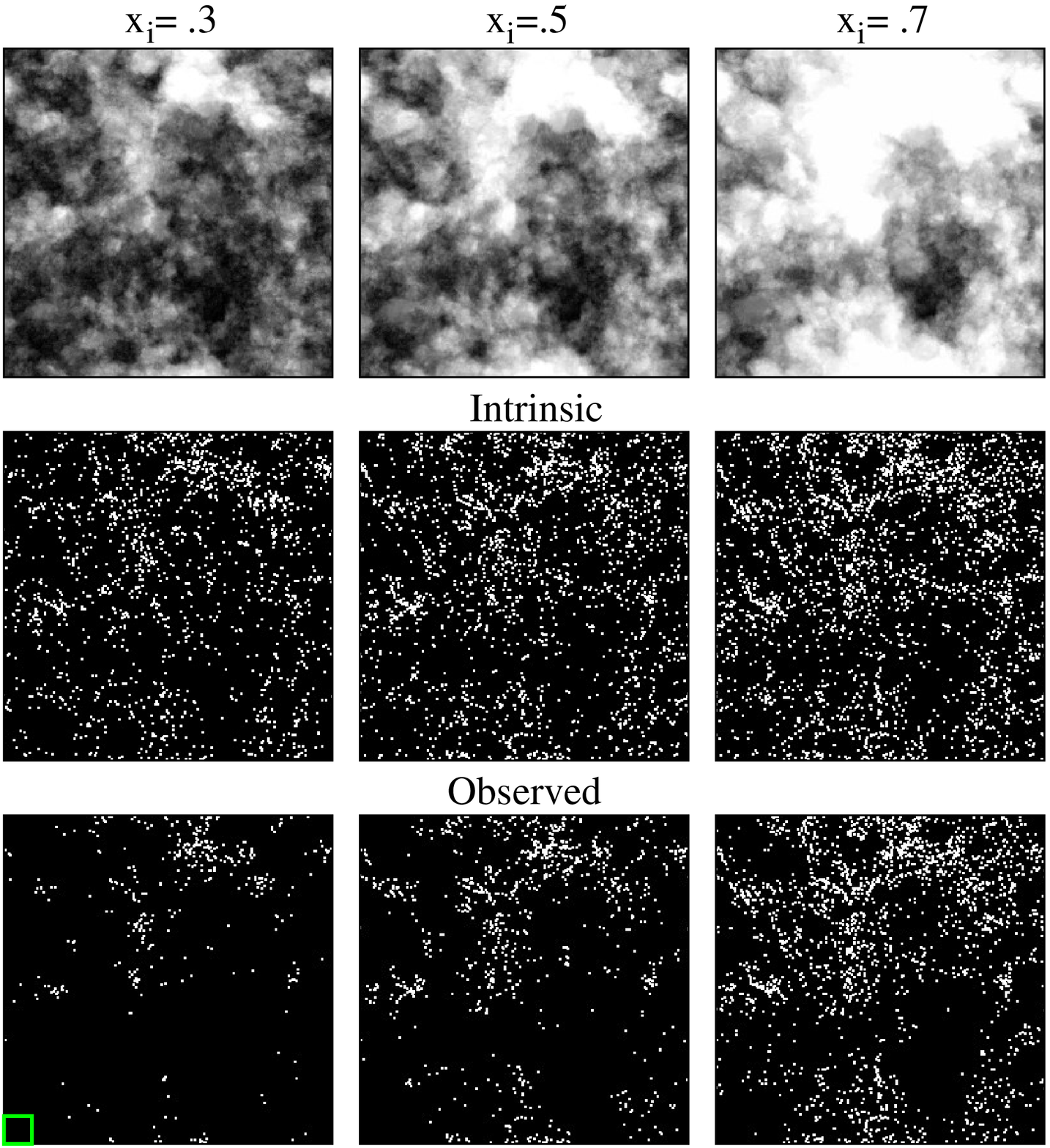}
\caption{ Simulation results (from McQuinn et al. 2007a\cite{MLya}) for the
relative distribution of neutral hydrogen and Ly$\alpha$ emitting galaxies
(LAE).  The top panels show the projection of the ionized hydrogen fraction
$x_i$ in the survey volume.  In the white regions the projection is fully
ionized and in black it is neutral.  The left, middle, and right panels are
for $z=8.2$ (with an average ionized fraction $\bar{x}_i = 0.3$), $z =7.7$
($\bar{x}_i= 0.5$), and $z = 7.3$ ($\bar{x}_i= 0.7$). The middle and bottom
rows are the intrinsic and observed LAE maps, respectively, for a
futuristic LAE survey that can detect halos down to a mass $> 1\times
10^{10} M_{\odot}$.  The observed distribution of emitters is modulated by
the location of the HII regions (compare bottom panels with corresponding
top panels).  Each panel is $94$ comoving Mpc across (or $0.6$ degrees on
the sky), roughly the area of the current Subaru Deep Field (SDF) at
$z=6.6$ \cite{Ka06}.  The depth of each panel is $\Delta
\lambda = 130$\AA, which matches the FWHM of the Subaru $9210$\AA
narrow-band filter. The number densities of LAEs for the panels in the
middle row are few times larger than the number density in the SDF
photometric sample of $z=6.6$ LAEs.  The large-scale modulation of LAE by
the HII bubbles is clearly apparent in this survey.  The square in the
lower left-hand panel represents the $3'\times 3'$ field-of-view of {\it
JWST} drawn to scale.
\label{LAE}}
\end{figure}

Given that the earliest galaxies created the ionized bubbles around them by
their UV emission, the locations of galaxies should correlate with the
cavities in the neutral hydrogen during reionization. Within a decade it
should be possible to explore the environmental influence of individual
galaxies by using large-aperture infrared telescopes in combination with
21-cm observatories of reionization \cite{WL07,WL05,WL07c}.

Wyithe \& Loeb (2007) \cite{WL07} calculated the expected anti-correlation
between the distribution of galaxies and the intergalactic 21cm emission at
high redshifts.  As already mentioned, overdense regions are expected to be
ionized first as a result of their biased galaxy formation.  This early
phase leads to an anti-correlation between the 21cm emission and the
overdensity in galaxies, matter, or neutral hydrogen. Existing Ly$\alpha$
surveys probe galaxies that are highly clustered in overdense regions. By
comparing 21cm emission from regions near observed galaxies to those away
from observed galaxies, future observations will be able to test this
generic prediction and calibrate the ionizing luminosity of high-redshift
galaxies. McQuinn et al. (2007a) \cite{MLya} showed that observations of
high-redshift Ly-alpha emitters (LAEs) have the potential to provide
definitive evidence for reionization because their Ly$\alpha$ line is
damped by the neutral regions in the IGM.  In particular, the clustering of
the emitters is increased by incomplete reionization (Fig. \ref{LAE}).  For
stellar reionization scenarios, the angular correlation function of the 58
LAEs in the Subaru Deep Field $z=6.6$ photometric sample
\cite{Taniguchi,Ka06} is already consistent with a mostly ionized IGM
\cite{MLya}. At higher redshifts near the beginning of reionization, when
the ionized regions were small, this analysis needs to be combined with
detailed radiative transfer calculations of the Ly$\alpha$ line, since the
line shape is sensitive to the local infall/outflow profile of the gas
around individual galaxies \cite{DL}.

\subsection{Gamma-ray Bursts: Probing the First Stars One
Star at a Time}
 
\begin{figure}
\centering
\includegraphics[height=8cm]{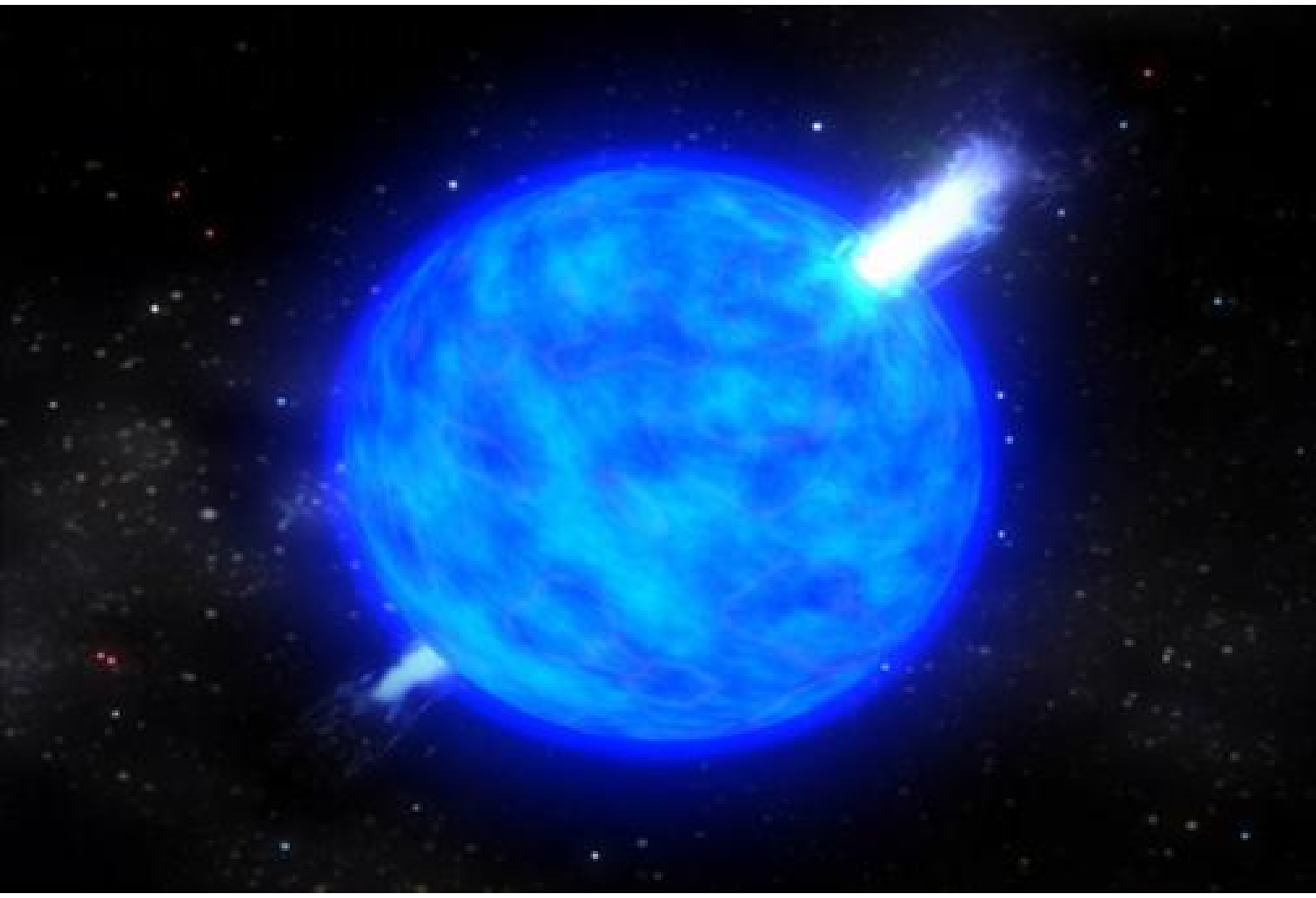}
\caption{Illustration of a long-duration gamma-ray burst in the popular
``collapsar'' model \cite{Zhang}.  The collapse of the core of a massive
star (which lost its hydrogen envelope) to a black hole generates two
opposite jets moving out at a speed close to the speed of light. The jets
drill a hole in the star and shine brightly towards an observer who happens
to be located within with the collimation cones of the jets. The jets
emanating from a single massive star are so bright that they can be seen
across the Universe out to the epoch when the first stars formed. Upcoming
observations by the {\it Swift}\/ satellite will have the sensitivity to
reveal whether Pop~III stars served as progenitors of gamma-ray bursts (for
more information see http://swift.gsfc.nasa.gov/).}
\label{grb}
\end{figure}

Gamma-Ray Bursts (GRBs) are believed to originate in compact remnants
(neutron stars or black holes) of massive stars. Their high luminosities
make them detectable out to the edge of the visible Universe
\cite{Lamb,CL00}. GRBs offer the opportunity to detect the most distant
(and hence earliest) population of massive stars, the so-called
Population~III (or Pop~III), one star at a time (Figure~\ref{grb}). In the
hierarchical assembly process of halos that are dominated by cold dark
matter, the first galaxies should have had lower masses (and lower stellar
luminosities) than their more recent counterparts. Consequently, the
characteristic luminosity of galaxies or quasars is expected to decline
with increasing redshift. GRB afterglows, which already produce a peak flux
comparable to that of quasars or starburst galaxies at $z\sim 1$--$2$, are
therefore expected to outshine any competing source at the highest
redshifts, when the first dwarf galaxies formed in the Universe.

\begin{figure}
\centering
\includegraphics[width=4in]{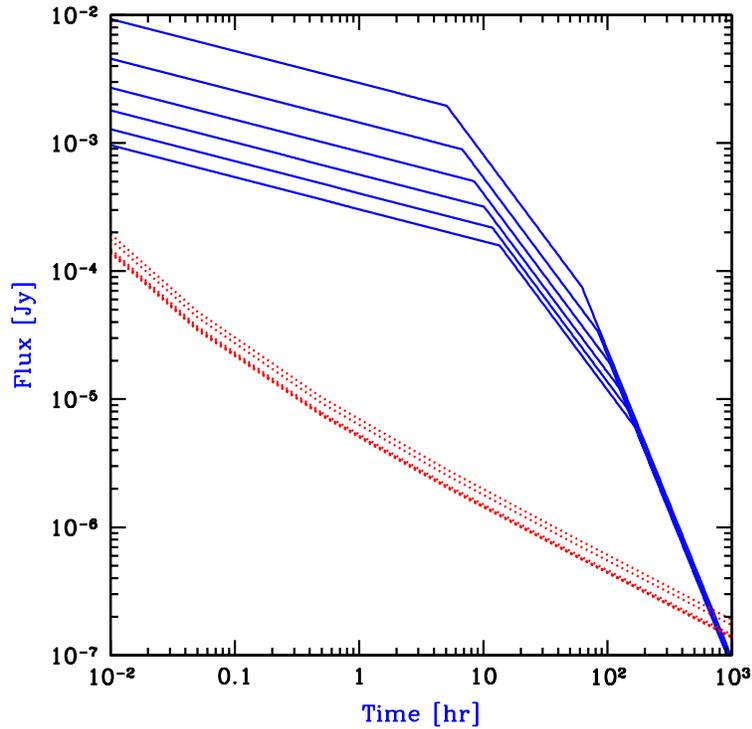}
\caption{GRB afterglow flux as a function of time since the $\gamma$-ray
trigger in the observer frame (from Barkana \& Loeb 2004a
\cite{BL04a}). The flux (solid curves) is calculated at the redshifted
Ly$\alpha$ wavelength. The dotted curves show the planned detection
threshold for the {\it James Webb Space Telescope} ({\it JWST}), assuming a
spectral resolution $R=5000$ with the near infrared spectrometer, a signal
to noise ratio of 5 per spectral resolution element, and an exposure time
equal to $20\%$ of the time since the GRB explosion. Each set of curves
shows a sequence of redshifts, namely $z=5$, 7, 9, 11, 13, and 15,
respectively, from top to bottom.}
\label{fig:GRB}
\end{figure}

GRBs, the electromagnetically-brightest explosions in the Universe, should
be detectable out to redshifts $z>10$. High-redshift GRBs can be identified
through infrared photometry, based on the Ly$\alpha$ break induced by
absorption of their spectrum at wavelengths below $1.216\, \mu {\rm m}\,
[(1+z)/10]$. Follow-up spectroscopy of high-redshift candidates can then be
performed on a large aperture infrared telescope, such as {\it JWST}. GRB
afterglows offer the opportunity to detect stars as well as to probe the
metal enrichment level \cite{FL03} of the intervening IGM. Recently, the
ongoing {\it Swift} mission \cite{Ge04} has detected GRB050904 originating
at $z\simeq 6.3$ \cite{Haislip}, thus demonstrating the viability of GRBs
as probes of the early Universe.

Another advantage of GRBs is that the GRB afterglow flux at a given
observed time lag after the $\gamma$-ray trigger is not expected to fade
significantly with increasing redshift, since higher redshifts translate to
earlier times in the source frame, during which the afterglow is
intrinsically brighter \cite{CL00}. For standard afterglow lightcurves and
spectra, the increase in the luminosity distance with redshift is
compensated by this {\it cosmological time-stretching} effect
\cite{CL00,BL04a} as shown in Figure~\ref{fig:GRB}.

GRB afterglows have smooth (broken power-law) continuum spectra unlike
quasars which show strong spectral features (such as broad emission lines
or the so-called ``blue bump'') that complicate the extraction of IGM
absorption features. In particular, the continuum extrapolation into the
Ly$\alpha$ damping wing during the epoch of reionization is much more
straightforward for the smooth UV spectra of GRB afterglows than for
quasars with an underlying broad Ly$\alpha$ emission line
\cite{BL04a}. However, the interpretation regarding the neutral fraction of
the IGM may be complicated by the presence of damped Ly$\alpha$ absorption
by dense neutral hydrogen in the immediate environment of the GRB within
its host galaxy \cite{BL04a,Totani} and by the patchiness of the neutral
IGM during reionization \cite{MGRB}. Since long-duration GRBs originate
from the dense environment of active star formation, the associated damped
Ly$\alpha$ absorption from their host galaxy was so-far always observed
\cite{Pro,Bloom}, including in the most distant GRB at $z=6.3$
\cite{Totani}.

\begin{acknowledgement}

I thank my collaborators on the work described in this review, Dan Babich,
Rennan Barkana, Volker Bromm, Benedetta Ciardi, Mark Dijkstra, Richard
Ellis, Steve Furlanetto, Zoltan Haiman, Jonathan Pritchard, George Rybicki,
Dan Stark, Stuart Wyithe and Matias Zaldarriaga.  I also thank Matt McQuinn
for a careful reading of the manuscript.

\end{acknowledgement}
%

%
%
%

\end{document}